\documentclass[reprint,NumberedRefs]{JASAnew}
\usepackage[caption=false]{subfig}
\usepackage{graphicx, amsmath, enumerate, natbib}
\usepackage{verbatim}
\usepackage{epstopdf}
\usepackage{ifthen}

\newboolean{singleColumn}
\setboolean{singleColumn}{false}

\DeclareMathOperator{\sinc}{sinc}

\DeclareMathOperator{\PAPR}{PAPR}

\begin{document}

\title{An Experimental Evaluation of the Generalized Sinusoidal Frequency Modulated Waveform for Active Sonar Systems}

\author{David A. Hague}
\email{david.a.hague@navy.mil}
\affiliation{Sensors and Sonar Systems Department, Naval Undersea Warfare Center, Newport, RI 02840}
\author{John R.~Buck} 
\email{johnbuck@ieee.org}
\affiliation{Department of Electrical and Computer Engineering, University of Massachusetts Dartmouth, Dartmouth, Massachusetts 02747}
\date{\today}

\begin{abstract}
This paper experimentally evaluates the Generalized Sinusoidal Frequency Modulated (GSFM) waveform, a generalization of the Sinusoidal Frequency Modulated (SFM) waveform.   The Instantaneous Frequency (IF) of the GSFM resembles the time/voltage characteristic of a Linear FM (LFM) chirp waveform.  Consequently, the GSFM possesses an Ambiguity Function (AF) that resembles a thumbtack shape.  Practical sonar system design must consider two factors beyond the AF.  The spectral efficiency (SE), defined as the ratio of energy in an operational frequency band to the total waveform energy, is another important metric for waveform design.  The Peak-to-Average-Power Ratio (PAPR) quantifies how close the waveform is to constant amplitude.  These measures predict a waveform's energy efficiency and ability to be accurately replicated on practical piezoelectric transducers, which have limits on both their bandwidth and maximum transmit power.  This paper explores these design considerations for the GSFM waveform and evaluates its performance against a host of other well established waveforms using simulated and experimental acoustic data.  The GSFM possesses superior SE, PAPR, and overall energy efficiency when compared to thumbtack waveforms.
\end{abstract}



\maketitle

\section{Introduction}
\label{sec:intro}

This paper experimentally evaluates the performance of the Generalized Sinusoidal Frequency Modulated (GSFM) waveform, a novel FM transmit waveform for active sonar \cite{HagueI, HagueII}.  Unlike the SFM whose Instantaneous Frequency (IF) function is a constant frequency sine wave, the GSFM modifies its IF function to resemble the time/voltage characteristic of a Linear FM (LFM) chirp waveform.  As a result of this modification, the GSFM achieves much lower range sidelobes than the SFM while preserving the excellent Doppler sensitivity properties of the SFM resulting in a thumbtack Ambiguity Function (AF).  Previous work \cite{HagueI, HagueII} established closed-form expressions for the GSFM waveform, its   AF, and compared its AF shape to other well established thumbtack waveforms.  The GSFM's AF performance is competitive with other established thumbtack waveforms, specifically for Time-Bandwidth Product (TBP) values less than 125.  However, sonar waveform design does not focus solely on AF shape.  Several practical issues impact the quality of the waveform transmitted by piezoelectric transducers, the most common transmit and receive devices employed by active sonar systems.  The system designer should ensure the waveform's target resolution properties (i.e., AF shape) are largely retained in the transmitted acoustic signal.  Additionally, the designer is concerned with maximizing the Sound Pressure Level (SPL) and therefore total energy of the finite duration transmitted acoustic signal which results in a stronger echo from a target.  The stronger echo signal will in turn possess a higher Signal to Noise Ratio (SNR) which improves detection performance in ambient noise-limited environments.  The properties that help achieve this are the waveform's Spectral Efficiency (SE) and Peak-to-Average Power Ratio (PAPR).   These practical considerations are interrelated and raise the question of whether the GSFM is better suited for transmission on piezoelectric devices compared to other established thumbtack waveforms.  This paper addresses this question by evaluating the GSFM's energy efficiency and the robustness of its AF shape using simulated and experimental acoustic data.  

A transducer is generally a resonant device whose frequency response drops off steadily away from the resonance frequency.  The phase of the transducer's frequency response is a non-linear function of frequency and the resulting group-delay of the transducer's frequency response is non-constant function of frequency.  When an FM waveform is transmitted or received by a transducer, each frequency component of the waveform  off resonance is attenuated in magnitude and shifted in phase.   The resulting transmitted FM acoustic signal therefore contains Amplitude Modulation (AM) and Phase Modulation (PM).  The electronics driving the transducer have a frequency response with a finite bandwidth and will also distort the transmitted acoustic signal.  In order to ensure the transmitted acoustic signal retains the waveform's mathematical properties, one typically designs a waveform to occupy the band of frequencies where the transducer is nominally flat, i.e., the region of the transducer's frequency response where the magnitude is nearly constant and the phase nearly linear.  It is therefore desirable to design a waveform that contains all or the vast majority of its energy in this operational band of frequencies.  Carson's bandwidth rule \cite{Couch} provides an approximate estimate of the bandwidth concentrating 98$\%$ of the waveform's energy.  However, a more accurate measure of this property is SE which directly computes the fraction of waveform energy contained in a specified band of frequencies.  This SE computation allows for a fair comparison among a set of waveforms.

Waveforms are typically tapered in time, specifically at the beginning and end of the waveform's duration, to slowly ramp up the waveform's amplitude to minimize the impact of transient effects from the transducer.  The tapering can additionally improve the waveform's SE.  However, this signal conditioning comes at a price.  The electronics driving the transducer are peak power limited and operating beyond this peak power limit can damage the electronics and introduce nonlinear distortions in the transmitted acoustic signal.  Even without the peak power constraint, a sonar system cannot transmit at arbitrarily high source levels without inducing cavitation.  To maximize the energy of the transmitted acoustic signal without suffering these negative effects, the waveform's average power should be as close as possible to its peak power.  This can be accomplished by designing a waveform with as nearly a constant amplitude as possible.  The PAPR measures the degree to which a waveform's amplitude is constant.  The lower a waveform's PAPR is, the more constant its amplitude is.

A waveform's SE and PAPR, while providing measures of different performance characteristics of a waveform design, are inter-related and also have an influence on other design characteristics.  The shape of the AF in time-delay and Doppler are determined by the shape of the magnitude squared of the waveform's spectrum and the complex envelope in time respectively \cite{Ricker}.  Changing the waveform's spectral shape in turn modifies its SE.  Tapering the waveform to modify its amplitude increases the PAPR as well as changes the spectral shape further modifying the waveform's SE.  These considerations present a variety of tradeoffs that the system designer must balance in order to achieve the desired system performance.  This paper evaluates the GSFM's performance for these practical design considerations, analyzes the design tradeoffs of these performance characteristics, and compares the GSFM's performance to that of other well established active sonar waveforms.  The rest of this paper is organized as follows: Section \ref{sec:waveformModel} describes the waveform signal model, the SE, PAPR, and the AF.  Section \ref{sec:Waveforms} describes several commonly used waveforms and introduces the SFM and the GSFM waveforms.  Section \ref{sec:results} evaluates the performance of the GSFM and compares its performance to that of other well known waveforms.  Finally, Section \ref{sec:conclusion} presents the conclusions.

\section{Waveform Model and Performance Metrics}
\label{sec:waveformModel}
This section describes the waveform signal model.  This paper assumes the sonar system is monostatic (i.e., the transmitter and receiver are co-located) and that the target of interest is a point target undergoing constant velocity motion.  These assumptions greatly simplify the analysis of the waveforms.  The analysis can be extended to more complicated models as design criteria dictate.

\subsection{The Waveform Model}
\label{subsec:waveformModel}
The transmit waveform signal $s\left(t\right)$ is modeled as a complex analytic signal with pulse length $T$ defined either over the interval $0 \leq t \leq T$ or $-T/2 \leq t \leq T/2$ expressed as
\begin{equation}
s\left(t\right) = w\left(t\right)e^{j\psi\left(t\right)} = w\left(t\right)e^{j\varphi\left(t\right)}e^{j2\pi f_c t}
\label{eq:ComplexExpo}
\end{equation}  
where $f_c$ is the carrier frequency, $\psi\left(t\right)$ is the instantaneous phase of the waveform, $\varphi\left(t\right)$ is the phase modulation function of the waveform, and $w\left(t\right)$ is a real-valued and positive amplitude tapering function \cite{Ricker}.  Unless otherwise specified, the amplitude tapering function $w\left(t\right)$ is assumed to be a rectangular function with amplitude $1/\sqrt{T}$ which normalizes the waveform to unit energy.  Utilizing a rectangular taper function results in a waveform whose IF function does not possess and AM contributions and is solely determined by the modulation function and carrier term.  The IF function of the rectangular tapered waveform is expressed as 
\begin{equation}
f\left(t\right) = \dfrac{1}{2 \pi}\dfrac{\partial \psi \left( t\right)}{\partial t} = \dfrac{1}{2 \pi}\dfrac{\partial \varphi \left( t\right)}{\partial t} + f_c.
\end{equation}  
The signal that is actually transmitted into the medium is the real component of the complex analytic signal
\begin{equation}
x\left(t\right) = \Re\{s\left(t\right)\} = w\left(t\right)\cos\left(\varphi\left(t\right) + 2\pi f_c t\right).
\label{realSignal}
\end{equation} 

Two important performance metrics to consider when transmitting a waveform on a practical transducer are SE and PAPR.  For FM waveforms, Carson's bandwidth rule \cite{Couch} estimates that 98$\%$ of a FM waveform's energy resides in a bandwidth $B$ expressed as $B = 2\left(\Delta f/2 + B_m\right)$ where $\Delta f$ is the peak frequency deviation of the waveform (i.e., swept bandwidth) and $B_m$ is the highest frequency component of the waveform's IF function.  Similar rules exist for Frequency Shift Keying (FSK) and Phase Coded (PHC) waveforms \cite{Couch}.  In order to provide a more accurate quantitative measure of SE as a fair means of comparison between waveforms, this paper defines the SE function denoted as $\Phi\left(\Delta F\right)$ of a transmit waveform as the ratio of waveform energy in a specific band of frequencies centered on $f_c$ with bandwidth $\Delta F$ to the total energy of the waveform across all frequencies expressed as 
\ifthenelse {\boolean{singleColumn}}
{\begin{equation}
\Phi\left(\Delta F\right) = \dfrac{\int_{f_c-\Delta F/2}^{f_c+\Delta F/2}|S\left(f\right)|^2df}{\int_{-\infty}^{\infty}|S\left(f\right)|^2df} = \int_{f_c-\Delta F/2}^{f_c+\Delta F/2}|S\left(f\right)|^2df
\label{eq:psi}
\end{equation}}
{\begin{equation}
\Phi\left(\Delta F\right) = \dfrac{\int_{f_c-\Delta F/2}^{f_c+\Delta F/2}|S\left(f\right)|^2df}{\int_{-\infty}^{\infty}|S\left(f\right)|^2df} = \int_{f_c-\Delta F/2}^{f_c+\Delta F/2}|S\left(f\right)|^2df
\label{eq:psi}
\end{equation}
}
where the waveform's energy in the denominator is assumed to be unity.  

The waveform's PAPR measures the ratio of the peak power of the transmitted acoustic signal $x\left(t\right)$ to its average power expressed in dB as 
\begin{equation}
\PAPR = 10\log_{10}\Biggl\{\dfrac{\max_t\{|x\left(t\right)|^2\}}{\frac{1}{T}\int_0^T|x\left(t\right)|^2dt}\Biggr\}.
\label{PAPR}
\end{equation}
For a given peak power limit, the PAPR is a measure of the waveform's average power.  For waveforms with the same duration $T$, the PAPR provides a measure of the total energy in the waveform.  A low PAPR implies a high average power and therefore high total energy.  Increasing the PAPR therefore reduces the total energy of the waveform.  The minimum possible PAPR for a signal in the form of \eqref{eq:ComplexExpo} is 3.02 dB, achieved by a constant frequency tone, i.e., $w(t) = 1$, $\varphi(t) = 0$.    Any tapering of the waveform that might be employed to improve the SE will also increase the PAPR introducing a tradeoff between SE and PAPR.

\subsection{The Broadband Ambiguity Function}
\label{subsec:BAAF}
The most common receiver employed in sonar systems is the Matched Filter (MF), or correlation receiver, as it is the optimal receiver for signal detection in the presence of additive white Gaussian noise \cite{Ricker}.  The MF is matched exactly to the echo signal only when the target is stationary relative to the sonar platform.  Targets with non-zero radial velocity $v$ with respect to the sonar transmitter and receiver introduce a Doppler effect to the echo signal.  The Doppler effect compresses or expands the signal in the time domain when the target is closing or receding respectively.  The AF measures the response of the MF to its Doppler scaled versions and is defined as \cite{Ricker}
\begin{equation}
\chi\left(\tau, \eta\right) = \sqrt{\eta} \int_{-\infty}^{\infty}s\left(t\right)s^*\left(\eta \left(t+\tau \right) \right) dt
\label{eq:BAAF}
\end{equation}
which is a function of time-delay $\tau$ and Doppler scaling factor $\eta = \left(1+v/c\right)/ \left(1-v/c\right)$ where $c$ is the speed of sound in the medium.  Of particular interest is the magnitude-squared of the AF $|\chi\left(\tau, \eta\right)|^2$ which will also be referred to as the AF in contexts where it is clear the magnitude squared of the AF is implied.  This paper focuses on waveforms that possess a ``thumbtack-like'' AF.  These waveforms' AF has a distinct mainlobe at the origin whose width in range and velocity is inversely proportional to the bandwidth and pulse length, respectively.   The rest of the waveform's AF volume, which is bounded, is spread roughly uniformly in the range-velocity plane \cite{Ricker}.  

\section{Waveforms}
\label{sec:Waveforms}
This section describes several common transmit waveforms and their AF shapes.  This section also introduces the SFM/GSFM waveforms and describes their performance in detail.

\subsection{The Linear Frequency Modulated Waveform}
\label{subsec:LFM}
\begin{figure}[h]
\centering
\includegraphics[width=1.0\reprintcolumnwidth]{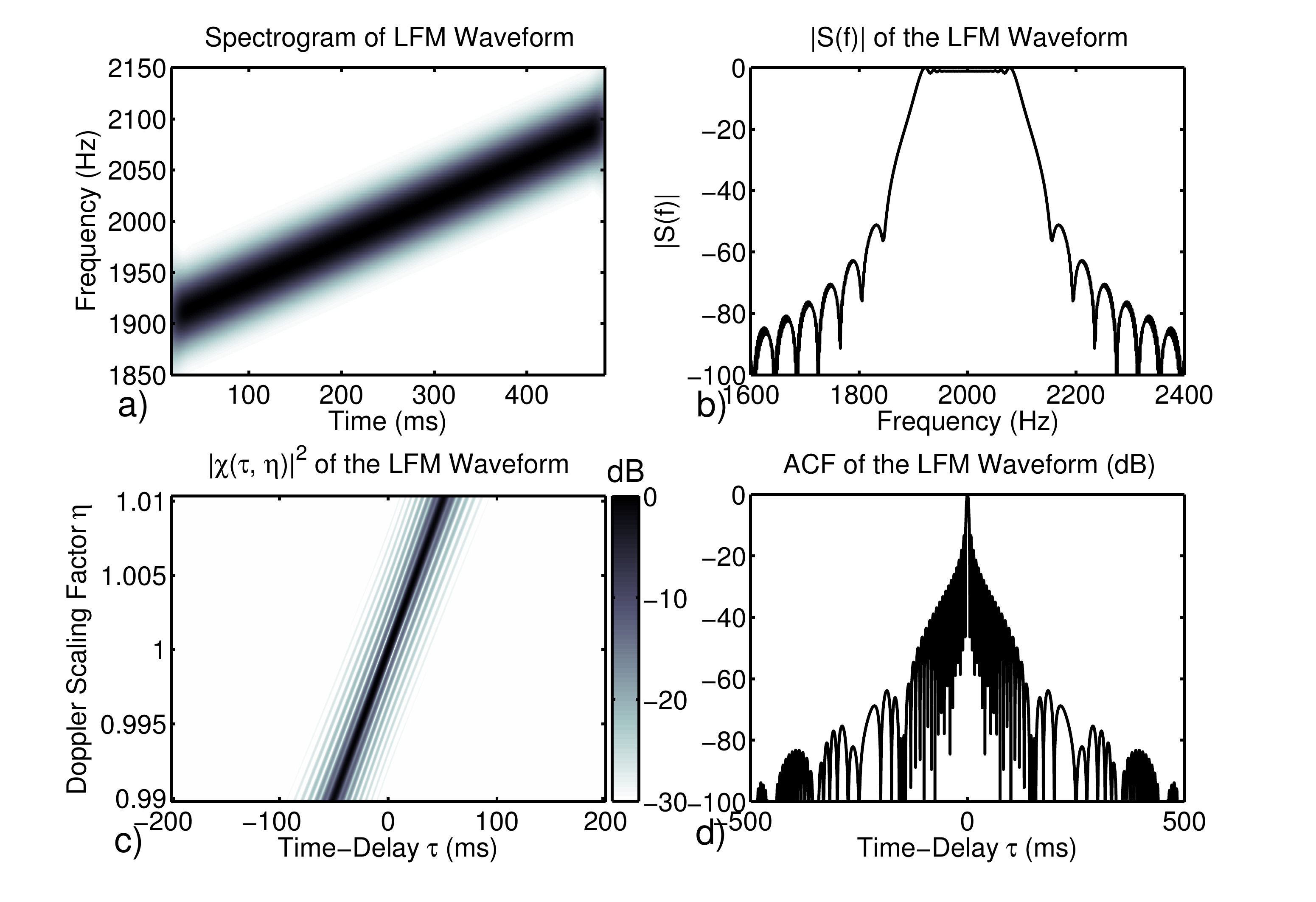}
\caption{\label{fig:LFM}Spectrogram (a), Spectrum (b), AF (c) ACF (d) of a LFM waveform with duration 0.5 s, $f_c$ = 2000 Hz, and swept bandwidth $\Delta f$ of 200 Hz.  By linearly sweeping through a band of frequencies $\Delta f$ throughout its duration, the LFM achieves the long duration and large bandwidth required for detecting and resolving closely spaced objects.}
\end{figure}

The LFM waveform is one of the most widely used sonar waveforms \cite{Ricker}.  The LFM achieves high range resolution by linearly sweeping across a band of frequencies $\Delta f$.  The LFM's phase and IF functions are expressed as 
\begin{equation}
\varphi_{LFM}\left(t\right) = \pi\left(\dfrac{\Delta f}{T}\right)t^2
\end{equation}
\begin{equation}
f_{LFM}\left(t\right) = \left(\dfrac{\Delta f}{T}\right)t + f_c
\end{equation}
for time $t$ defined as $-T/2 \leq t \leq T/2$.  Figure \ref{fig:LFM} shows the spectrogram, spectrum, AF, and ACF of a LFM waveform with duration 0.5 s, $f_c$ = 2000 Hz, and swept bandwidth $\Delta f$ of 200 Hz.  As seen from the spectrogram and spectrum, the LFM sweeps linearly across the band of frequencies $\Delta f$ and therefore places nearly equal energy across the swept band.  The LFM's AF has narrow mainlobe in time-delay whose width in inversely proportional the waveform's swept bandwidth $\Delta f$.  For non-zero target velocities, the AF's peak occurs at non-zero time-delays introducing a bias in the joint estimation of a target's range and velocity.  This bias, also known as range-Doppler coupling, limits the LFM to applications where range resolution is the system's main design goal.  When the LFM's fractional bandwidth is sufficiently small (i.e, $\leq 1/10$), the LFM is Doppler tolerant.  However, as the fractional bandwidth increases, the LFM becomes increasingly Doppler sensitive \cite{Lin}.  While not a thumbtack waveform, the LFM is a useful waveform to compare against due to its SE properties as well as its relative ease of implementation \cite{Ricker}.  Additionally, the LFM's roughly constant spectrum  is useful to determine the frequency response of a transducer.

\subsection{Thumbtack Waveforms}
\label{subsec:Thumtacks}

\begin{figure}[h]
\includegraphics[width=1.0\reprintcolumnwidth]{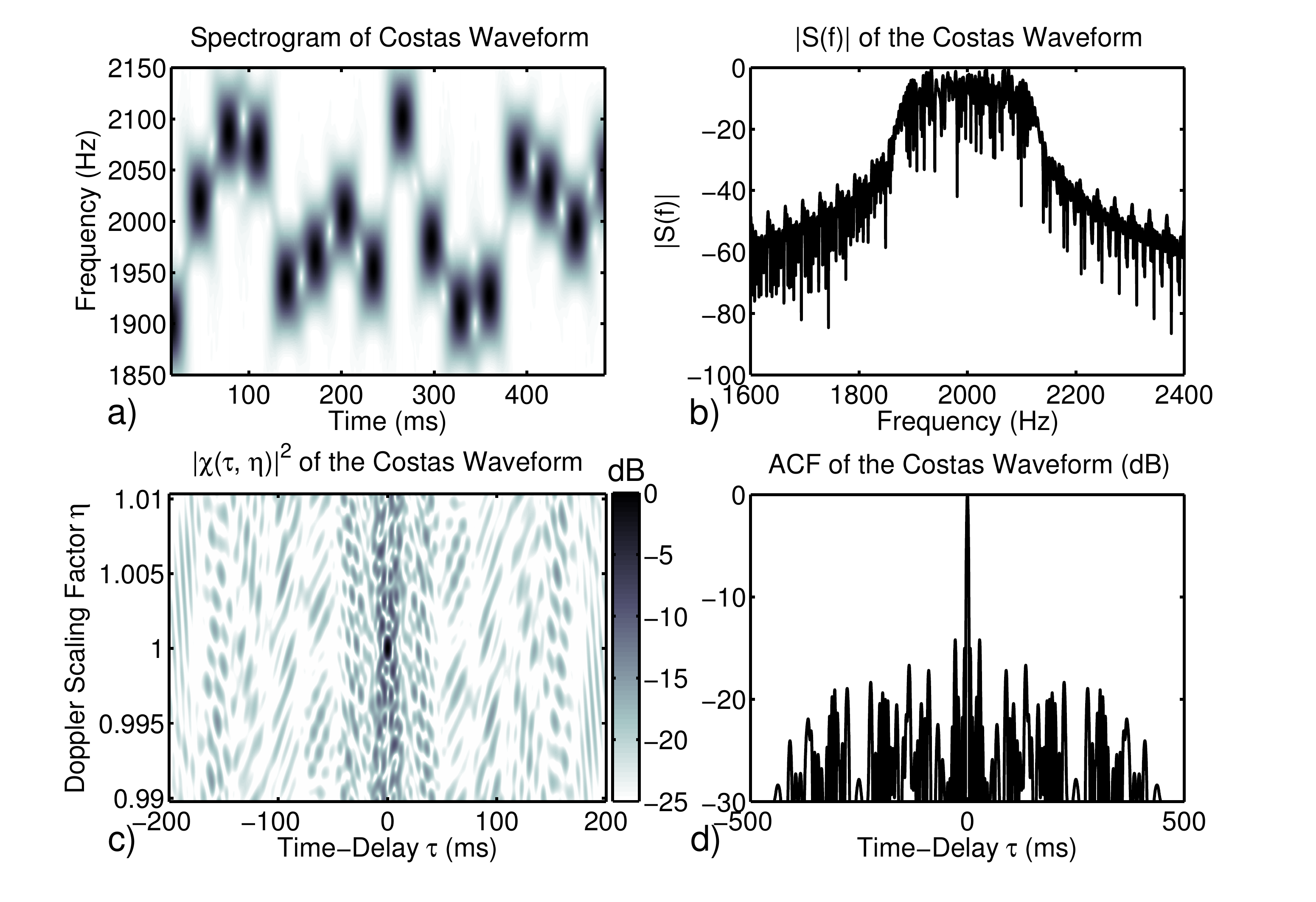}
\caption{Spectrogram (a), Spectrum (b), AF (c), and ACF (d) of a Costas waveform composed of 16 chips with duration 0.5 s, $f_c$ = 2 kHz, and a bandwidth $\Delta f$ of 200 Hz. }
\label{fig:Costas}
\end{figure}

This paper compares the GSFM waveform against three waveforms which attain a thumbtack AF; the Costas Waveform  \cite{Costas, PecknoldI}, Binary Phase-Shift Keying (BPSK) Waveform \cite{Ricker}, and Quadrature Phase-Shift Keying (QPSK) \cite{QuadPhase} waveform.  The Costas waveform of duration $T$ is a FSK waveform comprised of $N$ temporally contiguous pulses, called chips, each of which have duration $T/N$ and a frequency $f_i$ defined as
\begin{equation}
s_{c}\left(t\right) = \sum_{i=1}^{N} w\left(t - iT/N\right)e^{j \left(2 \pi f_i\left(t - iT/N\right) + \theta_i\right)}
\end{equation}
where $w\left(t\right)$ is the chip's amplitude tapering function and the phase term $\theta_i$ is included to ensure phase continuity between the chips in the waveform.  The frequency shift sequence $f_i$ for each chip is given by a Costas code \cite{Costas}.  Figure \ref{fig:Costas} shows the spectrogram, spectrum, AF, and ACF of a Costas waveform composed of 16 chips with duration 0.5 s, $f_c$ = 2 kHz, and a bandwidth $\Delta f$ of 200 Hz.  The Costas code minimizes the waveform's AF sidelobes and achieves a thumbtack AF.  However, the Costas waveform's spectral energy does not fall off as rapidly outside the frequency band as the LFM's.  The nearly instantaneous jump in frequency between the chips in the Costas waveform creates more out of band energy.  

\begin{figure}[h]
\includegraphics[width=1.0\reprintcolumnwidth]{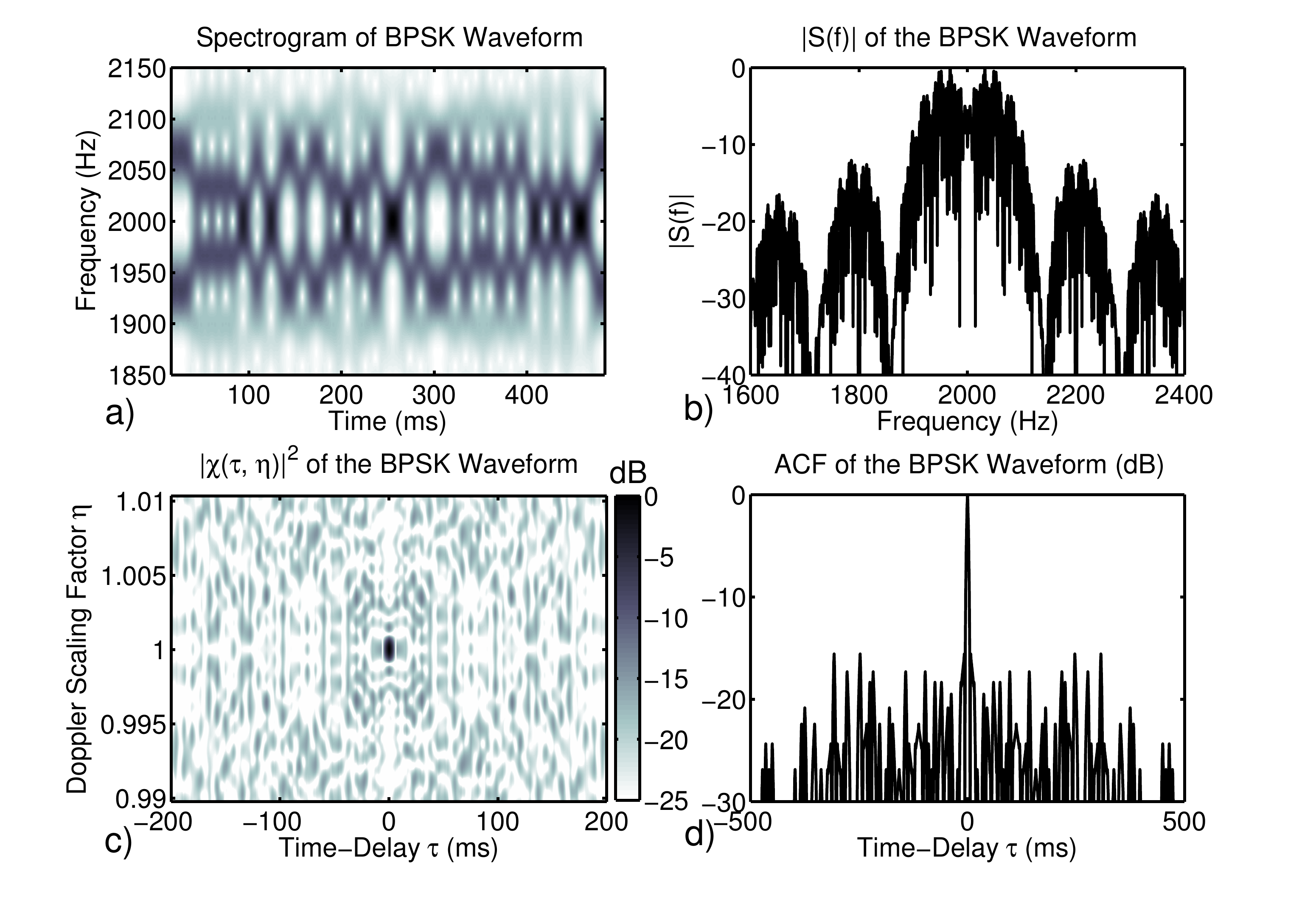}
\caption{Spectrogram (a), Spectrum (b), AF (c), and ACF (d) of a BPSK waveform composed of 64 chips with duration 0.5 s, $f_c$ = 2 kHz, and a bandwidth $\Delta f$ of 200 Hz. }
\label{fig:BPSK}
\end{figure}

The BPSK waveform differs from the Costas waveform by maintaining constant frequency in the chips, but changing the phase of each chip $\theta_i$ between 0 and $\pi$ according to a binary sequence.  The BPSK waveform is expressed as
\begin{equation}
s_{b}\left(t\right) = \sum_{i=1}^{N} w\left(t - iT/N\right)e^{j \left(2 \pi f_ct + \theta_i\right)}
\end{equation}
The phase sequence $\theta_i$ controls the AF shape of the BPSK waveform and a number of binary phase sequences have been designed to achieve desirable ACF properties \cite{Levanon, Li}.  Some of the most commonly used phase sequences are pseudo random sequences derived from Maximum Length Shift Register (MLSR) sequences \cite{Ricker} also known as m-sequences.  The resulting BPSK waveform is Doppler sensitive due its CW nature and the MLSR sequence helps spread the waveform's AF volume as evenly as possible resulting in a thumbtack AF.  Figure \ref{fig:BPSK} displays the spectrogram, spectrum, AF, and ACF of the BPSK waveform employing a m-sequence composed of 64 chips with duration 0.5 s, $f_c$ = 2 kHz, and a bandwidth $\Delta f$ of 200 Hz.  The BPSK's AF is not only thumbtack but even symmetric in time-delay (target range).  Much like the Costas waveform, the instantaneous jumps in phase between chips in the BPSK produces a waveform with less spectral containment than the LFM.  The BPSK waveform possesses a spectrum that is known as a ``noisy-sinc'' with spectral sidelobes falling off at 6 dB per octave.  Improving the SE of the BPSK requires signal conditioning to smooth the transitions in phase between the chips.  A simple way to achieve this is to apply an amplitude tapering function such as a Hann window to each chip.  The tapering will attenuate the signal to zero amplitude at the points in time where a phase transition occurs thus removing the transient at the phase transitions.  However the cost of such amplitude tapering is an increase the PAPR.  Another approach is to apply a form of Continuous Phase Modulation (CPM) to smooth the phase transitions between chips.  Perhaps the simplest CPM technique is the binary-to-quadriphase transformation proposed by Taylor and Blinchikoff \cite{QuadPhase} that produces a Quadriphase Shift Keying (QPSK) version of the original BPSK waveform.  

\begin{figure}[h]
\includegraphics[width=1.0\reprintcolumnwidth]{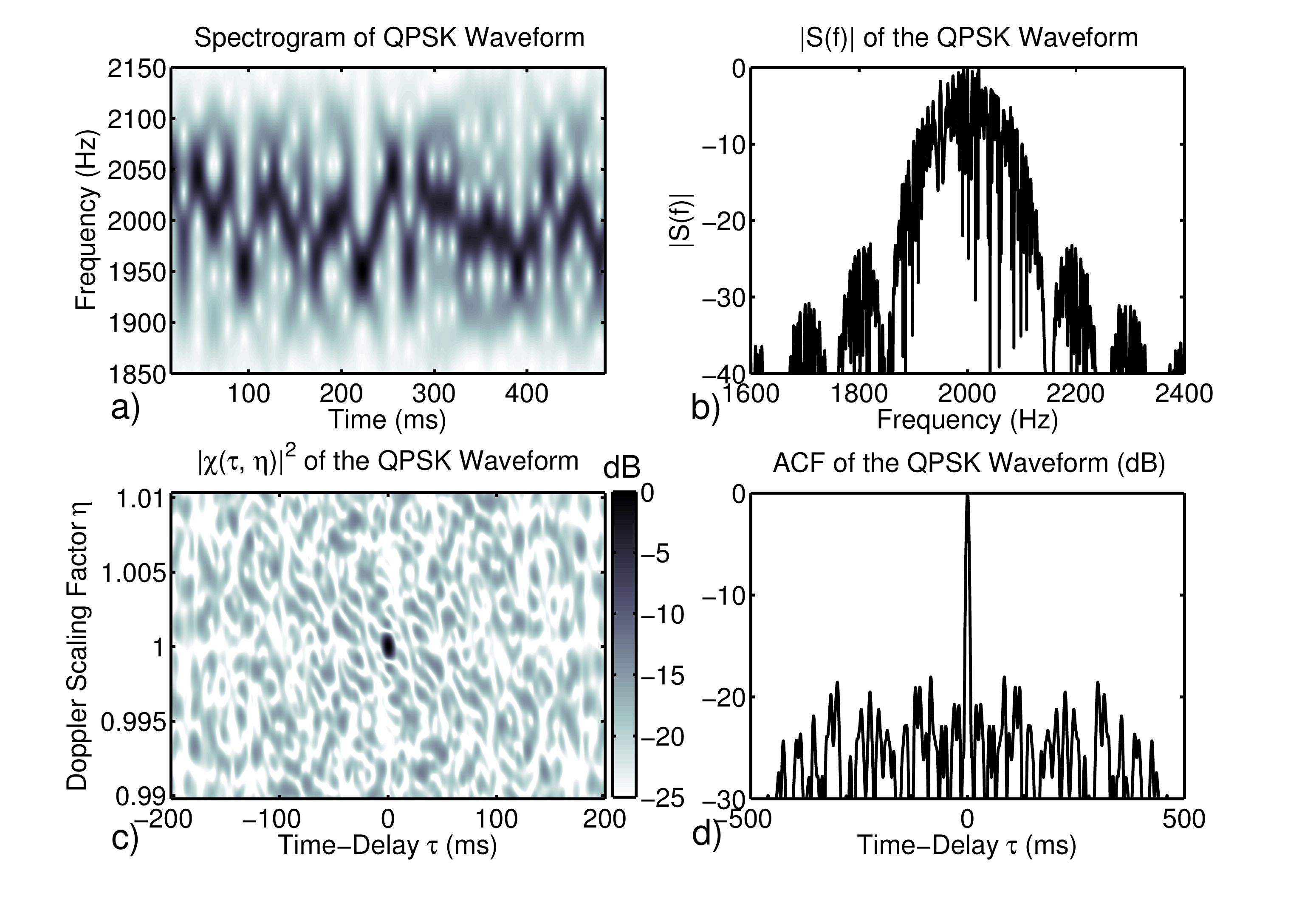}
\caption{Spectrogram (a), Spectrum (b), AF (c), and ACF (d) of a QPSK waveform composed of 100 chips with duration 0.5 s, $f_c$ = 2 kHz, and a bandwidth $\Delta f$ of 200 Hz. }
\label{fig:QPSK}
\end{figure}

The QPSK waveform results from a binary-to-quadriphase transformation which produces a waveform that maintains nearly the same AF shape as its binary phase counterpart, with reduced spectral sidelobes, and a nearly constant envelope.  The binary-to-quadriphase transformation is 
\begin{equation}
q_i = j^{\pm\left(i-1\right)}e^{j\theta_i}
\label{eq:biquad}
\end{equation}
where $\theta_i$ is the phase sequence.  Therefore, applying the transformation in \eqref{eq:biquad} to a MLSR sequence produces a thumbtack waveform with improved spectral efficiency over a BPSK and a constant envelope resulting in a low PAPR.  Figure \ref{fig:QPSK} shows the spectrogram, spectrum, AF and ACF of the QPSK created by performing the bi-phase to quad-phase transformation of the BPSK waveform shown in Figure \ref{fig:BPSK}.  The phase transitions between the chips of the QPSK waveform are linear which results in spectral sidelobes that fall off at 12 dB per octave.  As a result of this smoother phase transition, the QPSK possesses improved spectral containment over the BPSK waveform.  However, this signal conditioning comes at a cost.  As was noted in Ref\cite{Levanon}, modifying the phase of the QPSK also modifies the AF shape.  The AF mainlobe, while still thumbtack, is no longer even symmetric in time-delay, but sheared in time-delay and Doppler.  This shearing indicates the presence of delay-Doppler coupling, as seen in the LFM waveform.  The QPSK is an improvement in SE and PAPR over the BPSK waveform, but does not perfectly preserve the BPSK's AF properties.    

\subsection{The Sinusoidal Frequency Modulated (SFM) Waveform}
\label{SFM}
The SFM is a FM waveform whose IF function is a constant frequency sinusoid.  Its phase and IF functions are $\varphi_{SFM} \left( t \right) = \beta \sin \left(2 \pi f_m t\right)$ and
$f_{SFM} \left( t \right) = \beta f_m \cos \left(2 \pi f_m t \right)  + f_c$ where $\beta$ is the modulation index given as $\beta = \Delta f/2 f_m$, $f_m$ is the modulation frequency, and $\Delta f$ is the swept bandwidth \cite{Collins}.  There also exists the cosine phase counterpart of the SFM, the Cosine FM (CFM) whose instantaneous phase and frequency functions are shifted by $\pi/2$ radians.  The CFM maintains the same performance characteristics as the SFM and therefore this paper will focus solely on the SFM.  The spectrum of the SFM, derived in Appendix \ref{subsec:SFM_Spec}, is expressed as 
\begin{equation}
S_{SFM}\left(f\right) = \sqrt{T}\sum_{n=-\infty}^{\infty}J_n\{\beta\} \sinc\left[\pi T \left(f - f_c - f_m n \right) \right]
\label{eq:SFM_Spectrum}
\end{equation}
where $J_n\{\beta\}$ is the $n^{th}$ order cylindrical Bessel function of the first kind.
The expression in \eqref{eq:SFM_Spectrum} can be used to derive Carson's Bandwidth Rule \cite{Couch} for the SFM, $B_{SFM}$, expressed as 
\begin{equation}
B_{SFM} = 2\left(\beta + 1\right)f_m.
\label{eq:SFM_Carson}
\end{equation}  
Additionally, the SFM possesses a constant envelope.  In the absence of any amplitude tapering (i.e., a rectangularly windowed waveform) the SFM achieves the minimum PAPR of 3.02 dB.

\begin{figure}[]
\includegraphics[width=1.0\reprintcolumnwidth]{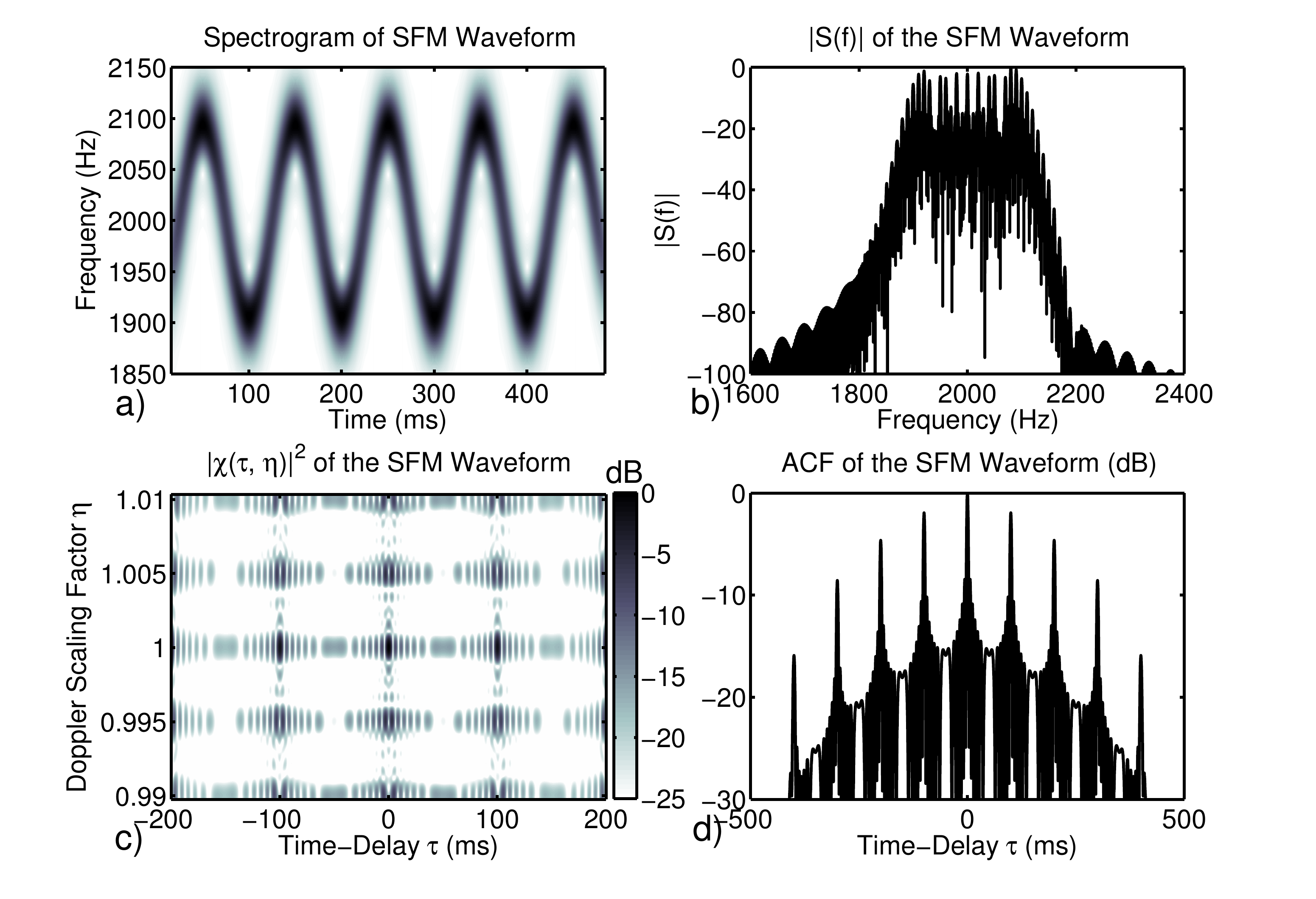}
\caption{Spectrogram (a), Spectrum (b), AF (c) and ACF (d) of a SFM with duration 0.5 s, $f_m$ = 10 Hz, $f_c$ = 2 kHz, and a swept bandwidth $\Delta f$ of 200 Hz.  The SFM possesses a comb spectrum and as a result is Doppler sensitive, but attains poor range resolution due to the high sidelobes in time-delay.}
\label{fig:SFM}
\end{figure}

The SFM's AF is \cite{HagueII}
\begin{multline}
|\chi\left(\tau, \eta\right)| \cong \dfrac{\sqrt{\eta}\left(T-|\tau|\right)}{T} \left|\sum_{n=- \infty}^{\infty}  J_n\{2 \beta \sin \left(\pi f_m \eta \tau \right) \} \times \right. 
\\ \left. \sinc \left[ \pi\left(\left(\eta-1\right)f_c -\dfrac{ f_m n \left(1+\eta \right)}{2} \right)\left(T-|\tau|\right) \right] \right|.
\label{eq:SFM_BAAF}
\end{multline}
Figure \ref{fig:SFM} shows the spectrogram, spectrum, AF, and ACF for an SFM of duration $T = 0.5$ s, a modulation frequency $f_m = 10$ Hz, a swept bandwidth $\Delta f = 200$ Hz, and a center frequency $f_c = 2$  kHz.  The SFM's IF function is clearly visible in the spectrogram and the SFM's spectrum is a "comb" function \cite{Newhall} with strong spectral lines equally spaced every $f_m$ Hz.  The AF is of the "bed of nails" variety with multiple high sidelobes in range and Doppler.  The SFM's AF possesses a distinct mainlobe whose widths in time-delay and velocity are inversely proportional to the waveform's bandwidth and pulse length, respectively.  The locations of the Doppler grating lobes are proportional to the modulation frequency $f_m$ and the range grating lobes are inversely proportional to $f_m$.  

The high sidelobes in range and Doppler of the SFM's AF are a direct result of the periodicity in the SFM's IF.  The periodic range sidelobes can be completely removed by designing an SFM with a single cycle in its IF, which is equivalent to reducing $f_m$ to $1/T$.  However, reducing the modulation frequency will  shift the high sidelobes in Doppler closer to the origin.  As a result, the high sidelobes may be located in a span of velocities where a realistic sonar target is expected.  These high sidelobes compromise the SFM's ability to resolve multiple targets in velocity.  Therefore, the SFM can be designed to estimate and resolve target range or target Doppler separately, but not both simultaneously.

\subsection{The GSFM}
\label{GSFM}
The GSFM is a FM waveform with phase and IF functions
\begin{equation}
\varphi_{GSFM} \left( t \right) = \left[\dfrac{\pi \Delta f}{\rho \left(2 \pi \alpha\right)^{\frac{1}{\rho}}} \right] S \{2 \pi \alpha t^\rho, 1/\rho   \}
\label{eq:GSFM_PHI}
\end{equation}
\begin{equation}
f_{GSFM} \left( t \right) = \left(\dfrac{\Delta f}{2}\right) \sin \left(2 \pi \alpha |t|^\rho\right)
\label{eq:GSFM_IF}
\end{equation}
where $S\{\}$ is the Generalized Sine Fresnel Integral \cite{NHMF}, $\rho$ is a variable exponent parameter that must be greater than or equal to 1, and $\alpha$ is a frequency modulation term with units $s^{-\rho}$.  Like the SFM, there are sine and cosine IF function versions of the GSFM.  Unlike the SFM however, the sine/cosine IF versions of the GSFM, can have noticeably different performance characteristics \cite{HagueDiss}.  This paper will focus solely on the GSFM with the sine IF function as shown in \eqref{eq:GSFM_PHI} and \eqref{eq:GSFM_IF}.

The spectrum of the GSFM, derived in Appendix \ref{subsec:GSFM_SPEC} is expressed as

\ifthenelse {\boolean{singleColumn}}
 {\begin{equation}
S_{GSFM}\left(f\right) = \sqrt{T}\sum_{n=-\infty}^{\infty} \mathcal{J}_n^{1:\infty}\{\beta_k\} \sinc\left[\pi T \left(f - f_c - \frac{a_0 \Delta f}{4} - \frac{n}{T} \right)\right],
\label{eq:GSFM_SpectrumI}
\end{equation}}
 {\begin{multline}
S_{GSFM}\left(f\right) = \sqrt{T}\sum_{n=-\infty}^{\infty} \mathcal{J}_n^{1:\infty}\{\beta_k\} \times \\  \sinc\left[\pi T \left(f - f_c - \frac{a_0 \Delta f}{4} - \frac{n}{T} \right)\right],
\label{eq:GSFM_SpectrumI}
\end{multline}}
where $\mathcal{J}_n^{1:\infty}\bigl\{\beta_k\bigr\}$ is the infinite dimensional Generalized Bessel Function (GBF) \cite{Dattoli} and $\beta_k$ are the Fourier coefficients of $\varphi\left(t\right)$.  Carson's Bandwidth Rule for the even-symmetric phase GSFM, derived in Appendix \ref{subsec:GSFM_Carson}, is expressed as 
\ifthenelse {\boolean{singleColumn}}
{\begin{equation}
B_{GSFM} = \left(\dfrac{\Delta f}{2\alpha\rho T^{\left(\rho-1\right)}}+1\right)
2\alpha\rho T^{\left(\rho-1\right)} = \Delta f + 2\alpha\rho T^{\left(\rho-1\right)}.
\label{eq:GSFM_Carson}
\end{equation}}
{\begin{equation}
B_{GSFM} = \Delta f + 2\alpha\rho T^{\left(\rho-1\right)}.
\label{eq:GSFM_Carson}
\end{equation}}
Also like the SFM, the GSFM possesses a constant evelope.  Appendix \ref{subsec:GSFM_PAPR} shows that the GSFM's PAPR is the minimum 3.02 dB.  The AF of the GSFM, derived in \cite{HagueII} is expressed as
\ifthenelse {\boolean{singleColumn}}
{\begin{multline}
\left|\chi \left(\tau, \eta\right)\right| \cong \dfrac{\sqrt{\eta}\left(T - |\tau|\right)}{T} \Biggl| \sum_{n=-\infty}^{\infty}\mathcal{J}_n^{1:N}\Biggl\{2\beta_k \sin\left(\frac{-\pi k \eta \tau}{T}\right); j^{-\left(k-1\right)} \Biggr\} \times \Biggr. \\ \Biggl. \sinc \left[\pi \left(T - |\tau|\right)\left( \left(\eta - 1\right) \left(f_c + \Delta f a_0/4\right) - \frac{\left(1+\eta\right) n}{2T} \right)\right] \Biggr|.
\label{eq:GSFM_BAAF}
\end{multline}}
{\begin{multline}
\left|\chi \left(\tau, \eta\right)\right| \cong \dfrac{\sqrt{\eta}\left(T - |\tau|\right)}{T} \\ \Biggl| \sum_{n=-\infty}^{\infty}\mathcal{J}_n^{1:N}\Biggl\{2\beta_k \sin\left(\frac{-\pi k \eta \tau}{T}\right); j^{-\left(k-1\right)} \Biggr\} \times \Biggr. \\ \Biggl. \sinc \left[\pi \left(T - |\tau|\right) \left( \left(\eta - 1\right) \left(f_c + \dfrac{\Delta f a_0}{4}\right) - \frac{\left(1+\eta\right) n}{2T} \right)\right] \Biggr|.
\label{eq:GSFM_BAAF}
\end{multline}}
Note that for $\rho = 1.0$, the GSFM reduces to the SFM, and the expressions \eqref{eq:SFM_Spectrum}-\eqref{eq:SFM_BAAF} are special cases of \eqref{eq:GSFM_SpectrumI}-\eqref{eq:GSFM_BAAF}.

\begin{figure}[!t]
\centering
\includegraphics[width=1.0\reprintcolumnwidth]{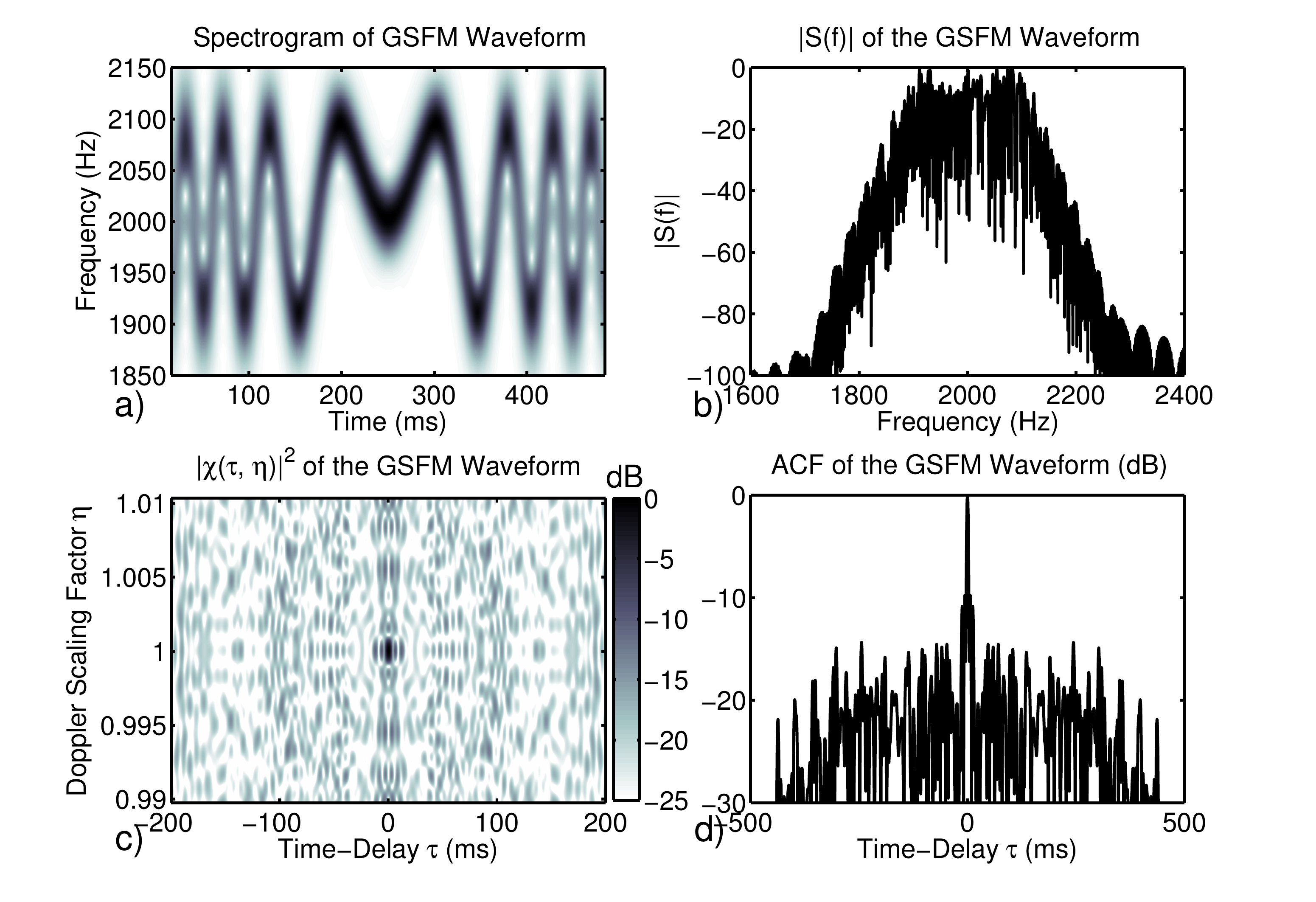}
\caption{Spectrogram (a), Spectrum (b), BAAF (c) Q-function (d) of a GSFM with duration 0.5 s, $\rho = 2.0$, $\alpha = 40$ s$^{-2}$, $f_c = 2$ kHz, and a swept bandwidth $\Delta f = 200$ Hz.  Because the IF of this variant of the GSFM has a time varying period, its AF possesses lower range sidelobes than the SFM while maintaining the Doppler sensitivity of the SFM.}
\label{fig:GSFM}
\end{figure}

The exponent parameter $\rho$ determines the overall shape of the IF function.  When $\rho = 2$ the resulting waveform's IF functions (\ref{eq:GSFM_PHI}) and (\ref{eq:GSFM_IF}) resemble the time/voltage characteristic of the LFM chirp waveform.  The LFM sinusoid IF variant of the GSFM does not exhibit the strict periodicity of the SFM's IF.  For any non-zero time-delay the spectral energy of the echo will not have substantial alignment with the IF of the MF replica resulting in a waveform with much lower ACF sidelobes than that of the SFM.  Defining the support interval as $0 \leq t \leq T$ generates a waveform whose IF function resembles the time-voltage characteristic of an up-sweeping chirp for $\rho > 1$.  This waveform has a non-symmetric IF.  Defining the support interval as $-T/2 \leq t \leq T/2$ and replacing the $t^{\rho}$ term with $|t|^{\rho}$ generates a waveform with an even-symmetric IF function that resembles the time-voltage characteristic of a base-banded chirp waveform.  The frequency modulation term $\alpha$ determines the number of cycles $C$ in the IF of the GSFM and is expressed as $C = \alpha T^{\rho}$ for a non-symmetric IF and $C = 2\alpha \left(T/2\right)^{\rho}$ for an even-symmetric IF.  Figure \ref{fig:GSFM} shows the IF function and AF of an even-symmetric GSFM with duration $0.5$ s, $\rho = 2.0$, $\alpha = 14$ s$^{-2}$ (or $C=7$), $f_c = 2$ kHz, and a swept bandwidth $\Delta f = 200$ Hz.  Unlike the SFM, this variant of the GSFM does not have a comb like spectrum and its AF exhibits a single distinct mainlobe centered at the origin with low to negligible sidelobes in range and Doppler while preserving the Doppler sensitivity of the SFM.  The GSFM's AF closely approximates a thumbtack AF, the original design goal of this work. 
  
\section{Simulations and Experimental Results}
\label{sec:results}

This section evaluates the GSFM's PAPR, SE, overall energy efficiency, and the robustness of its AF when transmitted on piezoelectric transducers using simulated and experimental data.  

\subsection{PAPR and Spectral Efficiency}
\label{subsec:PAPR}
As mentioned in the introduction, the SE $\Phi$ and PAPR of a waveform play important but competing roles in maximizing the total energy in the transmitted acoustic signal projected by a transducer.  Maximizing the energy of the acoustic signal will result in a stronger echo signal and therefore a higher received SNR.  In noise limited conditions, a higher SNR directly translates to improved probability of detection.  A system designer will typically employ transmit waveforms in a band of frequencies centered at the transducer's resonance frequency to maximize the source level and therefore total energy of the Transmitted Acoustic Signal (TAS).  Maximizing the concentration of the waveform's energy in this operational band of frequencies greatly aids in maximizing the energy in the TAS.  The SE metric measures the amount of waveform energy present in a specific band of frequencies and is computed by numerically evaluating \eqref{eq:psi}.  A higher SE implies a more spectrally contained waveform.  Maintaining constant amplitude in the waveform signal applied to the amplifier of the sonar transmitter additionally aids in maximizing the energy in the TAS.  The PAPR measures the peak to average power in the waveform signal applied to the amplifier of the sonar transmitter system.  The lower the PAPR of a waveform, the greater total energy it will contain.  

\begin{figure}[h]
\includegraphics[width=1.0\reprintcolumnwidth]{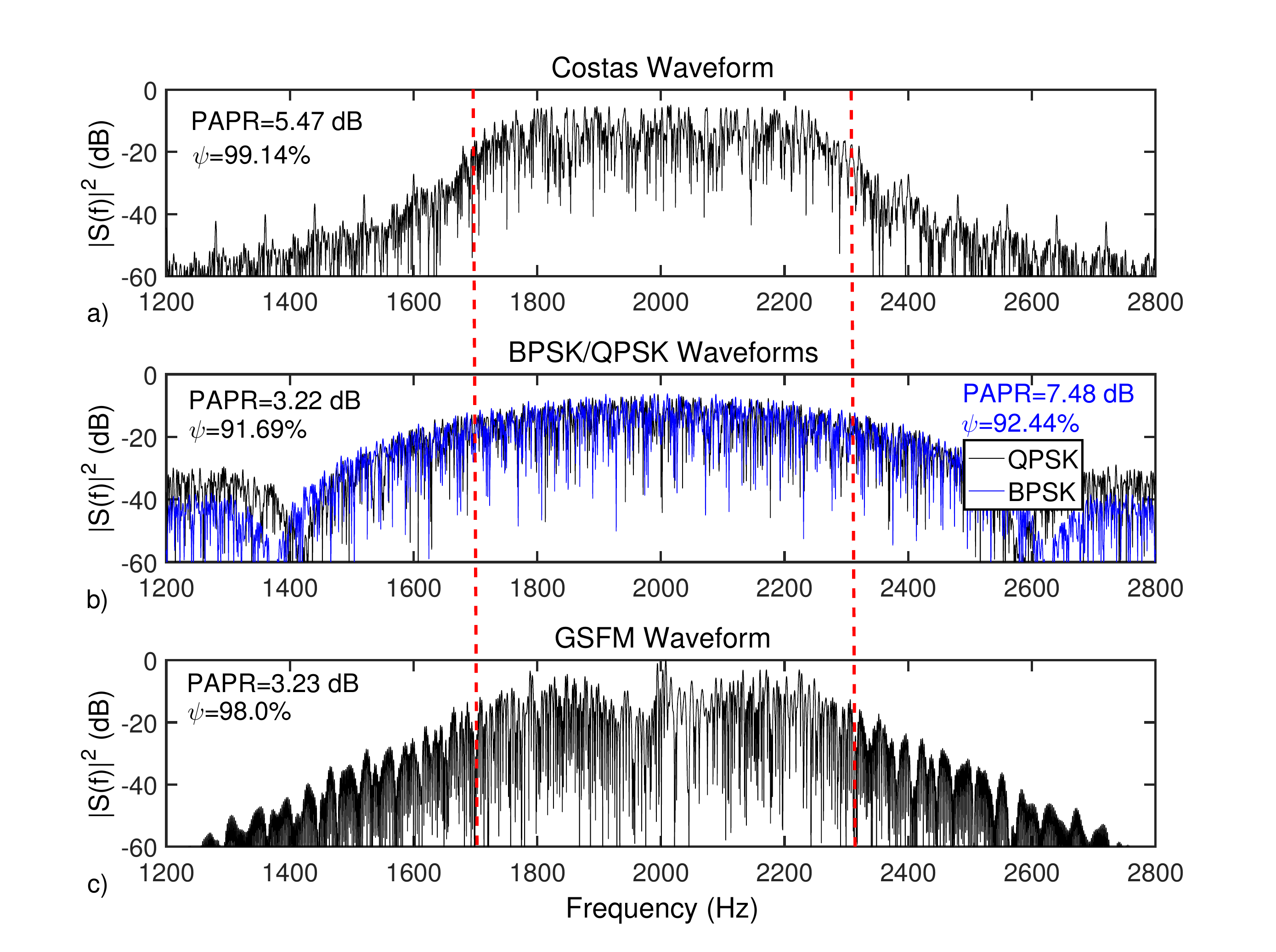}
\caption{(Color Online) SE of the Costas (a), BPSK and QPSK (b), and GSFM (c) waveforms.  The GSFM attains high SE while also requiring minimal tapering resulting in a low PAPR.  None of the Costas, BPSK, or QPSK waveforms can match the GSFM in SE or PAPR.}
\label{fig:SpecConI}
\end{figure}

Figure \ref{fig:SpecConI} shows the spectrum of the GSFM, Costas, BPSK, and QPSK waveforms all with duration $T = 0.5$ s, $f_c = 2000$ Hz.  The GSFM possesses a swept bandwidth $\Delta f = 500$ Hz and is tapered with a Tukey window \cite{Harris} with shape parameter $\alpha_T = 0.1$ with GSFM design parameters $\rho = 2.9$ and $\alpha = 253.78$ s$^{2.9}$.  The thumbtack waveforms were also tapered using a Tukey window to shape their respective spectra so that their ACF mainlobe widths matched the GSFM's.  This provided a fair comparison of each waveform's SE.  The Costas waveform chips were tapered with a Tukey window with shape parameter $\alpha_T = 0.85$ and the BPSK waveform chips were tapered with a Hann window \cite{Harris}.  In this particular example, the 98$\%$ bandwidth of the GSFM using \eqref{eq:psi} was numerically determined to be $632$ Hz, while using Carson's Bandwidth rule for the GSFM \eqref{eq:GSFM_Carson} over-estimates this same bandwidth by $65$ Hz, roughly a 10 $\%$ increase.  This was the general trend using \eqref{eq:GSFM_Carson} for GSFM waveforms with a variety of $\alpha$, $\rho$, and TBP values.  

The dashed lines in Figure \ref{fig:SpecConI} denote the band of frequencies centered about $f_c$ calculated by \eqref{eq:psi} for the GSFM.  The SE values for the other three waveforms were then computed using the same bandwidth $\Delta F$ of 632 Hz.  In this example, the Costas waveform actually has a slightly higher SE of $99\%$ than the GSFM but only achieves this SE due to tapering.  As a result of the tapering, the Costas waveform has a PAPR of $5.47$ dB, meaning the Costas waveform would transmit 2.24 dB less energy into the medium than the GSFM due to higher PAPR alone.  The BPSK's SE of $\Phi\left(\Delta f\right) = 92\%$ is notably less than the GSFM's even when tapering each chip with a Hann window.  The tapering resulted in the BPSK having a PAPR of $7.48$ dB.  In other words, the tapered BPSK would have transmitted 4.25 dB less energy than the GSFM.  An untapered BPSK with the same TBP would have a PAPR of $\approx 3.02$ dB and a SE of $\Phi\left(\Delta f\right) = 80\%$.  The QPSK waveform, whose PAPR of $3.22$ dB is roughly the same as the GSFM's, has a SE of $\Phi\left(\Delta f\right) = 92\%$, notably lower than the GSFM.  

\begin{figure}[h]
\includegraphics[width=1.0\reprintcolumnwidth]{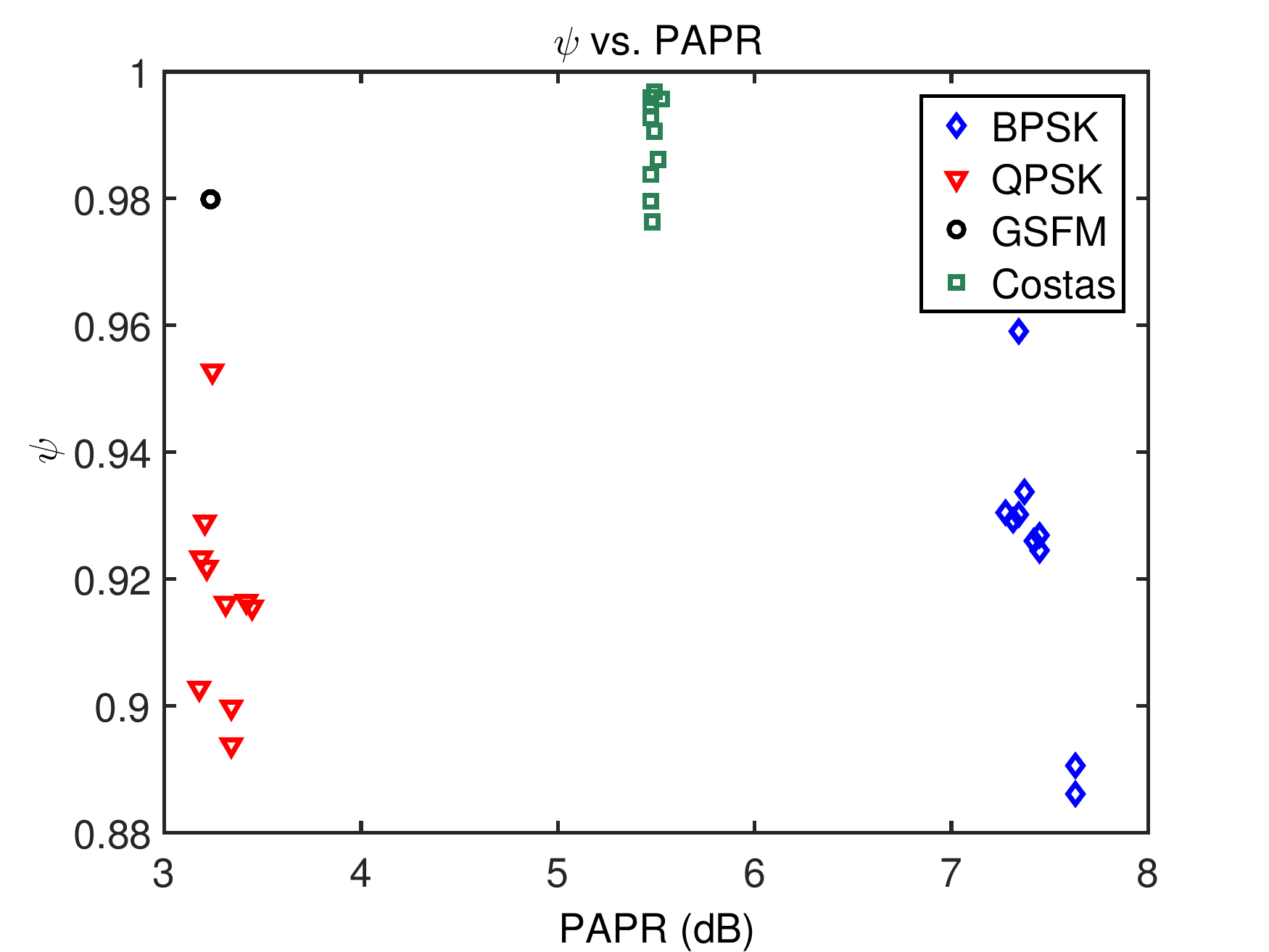}
\caption{(Color Online) SE of the GSFM, Costas, BPSK, and QPSK waveforms for TBPs 50, 75, 100, 125, 150, 175, 200, 250, 500, and 1000 .  For each TBP, the GSFM waveform attains high SE while also requiring minimal tapering resulting in a low PAPR.  None of the Costas, BPSK, or QPSK waveforms can match the GSFM in SE or PAPR for any of the TBPs tested.}
\label{fig:SpecConII}
\end{figure}

This is the general trend across TBPs as shown in Figure \ref{fig:SpecConII} which plots the SE $\Phi\left(\Delta F\right)$ as a function of PAPR for the GSFM, Costas, BPSK, and QPSK for TBPs $50, 75, 100, 125, 150, 175, 200, 250, 500,$ and $1000$ \cite{HagueI, HagueDiss}.  The 98$\%$ bandwidth $B$ was computed for the GSFM at each TBP using \eqref{eq:psi}.  The SE for the Costas, BPSK, and QPSK waveforms for each TBP were also computed with $\Delta F$ set to the GSFM's 98$\%$ bandwidth $B$ as a means to compare each waveforms SE at each TBP value.  Additionally, the PAPR was computed for all the waveforms.  The design goal is for a waveform to possess both high SE and a low PAPR which directly translates to data points that tend to the upper left corner of the figure.  The GSFM data points all appear in the same location ($\PAPR = 3.23$ dB, $\Phi\left(\Delta f\right) = 0.98$) and are closest to the upper left corner of Figure \ref{fig:SpecConII} meaning the GSFM attains high SE and a low PAPR.  None of the Costas, BPSK, or QPSK waveforms can match the same performance in both SE and PAPR of the GSFM for any of the TBPs tested.  The Costas waveforms match the SE of the GSFM, but fall short on PAPR.  Conversely, the QPSK waveforms match the PAPR of the GSFM, but fall short on SE.   Finally, the BPSK waveforms are inferior to the GSFM for both SE and PAPR.  While SE is an important design consideration, the following section demonstrates that PAPR has the greater impact on overall energy efficiency.  Considering SE alone, the GSFM would be considered only marginally more efficient than all the waveforms examined except the Costas waveform.  However, the low PAPR of the GSFM makes it several dB more energy efficient than the amplitude tapered Costas and BPSK waveforms.  When considering SE and PAPR jointly, the GSFM clearly stands out as the most energy efficient of the waveforms examined.      

\subsection{Transducer Replicas}
\label{subsec:Replica}
The results from the previous section show that the GSFM possesses both high SE and low PAPR, two important characteristics to consider when transmitting waveforms on piezoelectric transducers.  In addition to ensuring that a waveform has these desirable properties, a system designer will  record a replica of the acoustic signal that is transmitted and received by the system's transducers, referred to here as the Transducer Replica Waveform (TRW).  During system calibration, the system designer will verify that the TRW is an accurate representation of the transmit waveform that was designed in simulations and that the TRW maintains its AF properties.  In many cases the TRW will be used as the base MF from which other Doppler scaled MF's will be derived.  As mentioned earlier, transducers are resonant devices whose frequency response varies in magnitude and phase as function of frequency.  This introduces AM and PM distortion into the TRW, which can drastically change the AF shape of the TRW if the waveform's spectrum extends substantially above or below resonance.  

An equalizer filter can compensate for the transducer's frequency response and thereby minimize distortion in the TRW.  Applying an equalizer in a system with a peak power limit means attenuating the frequency components of the waveform that are at or near the resonance frequency.  This attenuation reduces the source level of the transmitted acoustic signal, reducing the echo strength which in turn reduces SNR and therefore detection performance.  The system designer is presented yet another design tradeoff: equalize the waveform's spectrum to accurately replicate the waveform on the device and therefore preserve its AF properties, or omit equalization and maximize the TRW's energy for detection while accepting degradation in resolution performance.  These tradeoffs motivated the TRW experiments described in this section.    

The waveform transmission experiments were conducted at the Naval Undersea Warfare Center's (NUWC) test tank facility.  An F-82 \cite{F82} projector transmitted the waveforms which were then received by a F-52 hydrophone. The devices were separated by 4.5 meters at a depth of 4.8 meters and operated between $100-120$ kHz.  The transmission system consisted of an arbitrary waveform generator with an amplification and impedance matching stage which generated the electronic waveform signal that drove the projector transducer.  The receive end of the system included of a lowpass filter with a cutoff frequency of 250 kHz followed by a 24-bit analog to digital converter with sampling frequency $f_s = 2 $ MHz.  The transmitted waveforms were of duration $T=5$ ms with center frequency $f_c = 110$ kHz and possesed a swept bandwidth $\Delta f$ of $10$ kHz (narrowband) or $20$ kHz (wideband).  The short duration of the waveforms ensured that a full replica could be recorded without receiving any echoes from within the test tank.  The transmit/receive components could be adjusted to equalize the frequency response of the entire system.  This option, when utilized, produced a frequency response that only varied by $0.39$ dB across the $100-120$ kHz band.  When not equalized, the overall transmit/receive system varied by $4.07$ dB across that same band.  Table I lists the waveforms transmitted for this test.  Both equalized and non-equalized TRWs were transmitted during this test.  It is important to note that the technicians at NUWC's test tank facility wanted to ensure that the transmitted waveforms had sufficient spectral containment to reduce the risk of damaging the transducer and driving electronics.  To address this concern, no QPSK or un-tapered BPSK waveforms were transmitted during this test.  All BPSK waveforms were tapered using a Hann window.    

\begin{table}[htb]

\begin{center}
{
\begin{tabular}{|c||c|c|c|}\hline
Waveform  & $\Delta f$ (kHz) & Taper & Properties \\\hline
LFM I       & 10  & Tukey $0.1$	 		 &  NA 	      \\\hline
LFM II      & 20  & Tukey $0.1$		 &  NA 		 \\\hline 
Costas I    & 20  & Tukey $0.1$		 & 10 Chips \\\hline
Costas II   & 20  & Hann     		      & 10 Chips \\\hline
BPSK I      & 10  & Hann     			 & 15 Chips \\\hline
BPSK II     & 20  & Hann     			 & 31 Chips \\\hline
GSFM I     & 10  & Tukey $0.1$	 	 & C=9.5, $\rho=2.2$ \\\hline
GSFM II    & 10  & Tukey $0.1$	 	 & C=7.5, $\rho=2.0$ \\\hline
GSFM III   & 20  & Tukey $0.1$	 	 & C=15, $\rho=2.55$ \\\hline
GSFM IV   & 20  & Tukey $0.1$	 	 & C=13, $\rho=2.3$ \\\hline

\end{tabular}
\caption[List of transmit waveforms evaluated.]{List of waveforms and their pertinent properties transmitted in NUWC test tank.}
}
\end{center}
\label{table:EEI}
\end{table}  

There were two main objectives for this test.  The first objective was to compare the AF of the equalized  and non-equalized TRW so as to assess the degree of distortion between the two TRW's AFs.  The second objective of this experiment is to compare the overall energy efficiency of each TRW.  The energy efficiency of the TRW $\tilde{E}$, compares the amount of energy in the GSFM's TRW $E_{GSFM}$ to the amount of energy in the LFM and thumbtack waveforms' TRW $E_w$.  This measure of energy efficiency is calculated using the formula
\begin{equation}
\tilde{E} = 10\log_{10}\left(\dfrac{E_w}{E_{GSFM}} \right).
\label{eq:energyEfficiency}
\end{equation} 
The energy of the TRW \eqref{eq:energyEfficiency} is a measure that combines the effects of waveform's SE, PAPR, and also accounts for the transducer's frequency response providing a comprehensive measure of a waveform's overall energy efficiency.

\begin{figure}[h]
\centering
\includegraphics[width=1.0\reprintcolumnwidth]{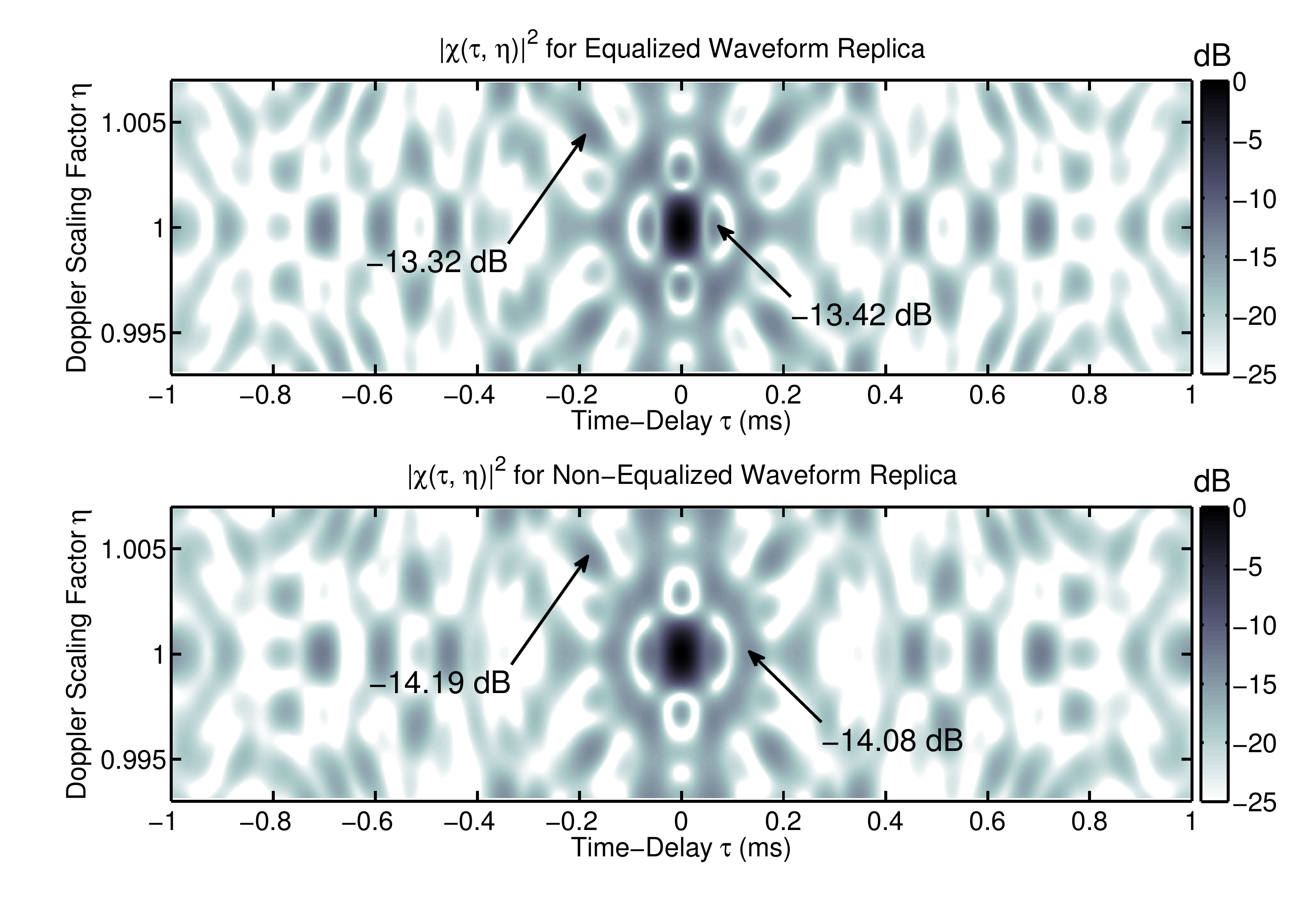}
\caption{AFs of the equalized and non-equalized TRW for GSFM IV.  Both TRWs AFs look nearly identical with the non-equalized TRW's AF having a slightly wider mainlobe width in time-delay and some sidelobe heights were reduced.}
\label{fig:RepComp}
\end{figure}

Figure \ref{fig:RepComp} shows the AF of the equalized and non-equalized GSFM IV TRW.  The mainlobe of the non-equalized TRW is slightly widened in time-delay.  An analysis of the time-delay mainlobe showed the non-equalized TRW's 3 dB mainlobe width was widened by approximately $10\%$. Visual inspection of the PSLs of the two TRWs' AFs shows that while most sidelobes have not substantially changed in value, some sidelobes were reduced in height.  Overall, the non-equalized TRW's AF shape is very similar to the equalized TRW with very little noticeable differences.  An analysis of the TRWs for all the evaluated waveforms in Table I displayed similar characteristics to the results shown in Figure \ref{fig:RepComp}.

Tables II and III show the energy efficiency $\tilde{E}$ for the narrowband and wideband TWRs respectively.  All the narrowband TRWs energy efficiency was referenced to the energy of non-equalized GSFM II TRW.  For the equalized TRWs in narrowband, only the LFM was marginally more efficient than the GSFMs and substantially more efficient than the Hann tapered BPSK.  For the non-equalized TRWs, the GSFM was more efficient.  The wideband TRWs were referenced to the non-equalized GSFM IV TRW.  Both the equalized and non-equalized wideband GSFM TRWs were the most energy efficient.  
 
\begin{table}[htb]

\begin{center}
{
\begin{tabular}{|c|c||c|c|}\hline
Eq Waveform        & $\tilde{E}$ (dB)  & Non-Eq Waveform   & $\tilde{E}$ (dB) \\\hline
LFM I       		      & $-1.09$ 			 & LFM I				 &  $-0.39$	      \\\hline
BPSK I      			& $-5.23$   		 & BPSK I				 &  $-4.36$      \\\hline 
GSFM I       		& $-1.26$ 			 & GSFM I  			 &  $-0.03$ 	 \\\hline
GSFM II     		& $-1.26$ 			 & GSFM II  	  		 &  $0.00$ 	      \\\hline
\end{tabular}
\caption[Evaluation of Energy Efficiency of the narrowband TRWs referenced to the total energy of GSFM II.]{Evaluation of Energy Efficiency of the narrowband TRWs referenced to the total energy of GSFM II.  For the equalized TRWs, the LFM is marginally more efficient than the GSFMs and substantially more efficient than the BPSK TRWs. For the non-equalized TRWs, the GSFMs possessed higher energy efficiency.}
}
\end{center}
\label{table:EEI_I}
\end{table}  

\begin{table}[htb]

\begin{center}
{
\begin{tabular}{|c|c||c|c|}\hline
Eq Waveform        & $\tilde{E}$ (dB)  & Non-Eq Waveform   & $\tilde{E}$ (dB) \\\hline
LFM II 	       		& $-0.99$	 		 & LFM II     	     		 &  $-0.14$	 \\\hline
Costas I		      & $-1.08$	 		 & Costas I    	  		 &  $-0.11$ 	 \\\hline
Costas II		      & $-5.07$	 		 & Costas II    	  		 &  $-4.08$ 	 \\\hline
BPSK II			      & $-5.16$	 		 & BPSK II    	  		 &  $-4.04$ 	 \\\hline
GSFM III              & $-0.97$	 		 & GSFM III  	  		 &  $-0.03$ 	 \\\hline
GSFM IV		      & $-0.96$	 		 & GSFM IV    	  		 &  $0.0$ 	 \\\hline
\end{tabular}
\caption[Evaluation of Energy Efficiency of the wideband TRWs referenced to the total energy of GSFM IV.]{Evaluation of Energy Efficiency of the wideband TRWs referenced to the total energy of GSFM IV.  For both equalized and non-equalized TRWs, the GSFMs were the most energy efficient.}
}
\end{center}
\label{table:EEI_II}
\end{table} 

\section{Conclusion}
\label{sec:conclusion}
This paper explored the GSFM's waveform's spectral containment, energy efficiency, and the fidelity of replicas transmitted on practical piezoelectric devices.  The GSFM's performance was compared to the LFM as well as the Costas, BPSK, and QPSK thumbtack waveforms.  The GSFM contains the vast majority of its energy in a confined band of frequencies while requiring minimal amplitude tapering which as a result achieved both higher SE and a lower PAPR than any of the thumbtack waveforms evaluated.  The test tank experiments showed that the GSFM's TRW largely maintains its desireable AF shape when transmitted on piezoelectric transducers and that its overall energy efficiency surpasses that of other known thumbtack waveforms.  The decision of whether to utilize the GSFM over some other established thumbtack waveform in a practical system, assuming the system's mission requires or benefits from a thumbtack waveform, would certainly consider resolution and AF properties.  Generally speaking, the GSFM is competitive with the other known thumbtack waveforms for lower TBPs (i.e. less than 125) \cite{HagueI, HagueDiss}.  However, the results described in this paper show that when considering energy efficiency, the GSFM outperforms other known thumbtack waveforms.  If the system designer weighs energy efficiency above all other considerations, then the energy efficient GSFM would be the better choice for a thumbtack waveform.

\appendix
\section{Derivation of the Spectrum for the SFM and GSFM}
\label{sec:SFMGSFMSpectrum}

\subsection{The SFM's Spectrum}
\label{subsec:SFM_Spec}
Using the waveform definition in \eqref{eq:ComplexExpo} over the interval $-T/2 \leq t \leq T/2$ with rectangular window tapering function with height $\frac{1}{\sqrt{T}}$ to ensure unit energy, the Fourier transform of the SFM waveform is expressed as
\begin{equation}
S_{SFM}\left(f\right) = \dfrac{1}{\sqrt{T}}\int_{-T/2}^{T/2}e^{j\beta \sin\left(2 \pi f_m t\right)}e^{j2\pi \left(f_c -f\right) t}dt
\end{equation}
Using the Jacobi-Anger expansion \cite{Abramowitz}, the expression is simplified to
\begin{equation}
S_{SFM}\left(f\right) = \dfrac{1}{\sqrt{T}}\sum_{n=-\infty}^{\infty}J_n\{\beta\} \int_{-T/2}^{T/2} e^{-j 2 \pi \left(f - f_c - f_m n \right)t}dt
\end{equation} 
Carrying out the integral and utilizing the frequency shift Fourier transform property yields the result
\begin{equation}
S_{SFM}\left(f\right) = \sqrt{T}\sum_{n=-\infty}^{\infty}J_n\{\beta\} \sinc\left[\pi T \left(f - f_c - f_m n \right) \right]
\label{eq:SFM_Spec_Der}
\end{equation}
This result differs from the results in modern communication theory textbooks \cite{Couch} in that the expression in \eqref{eq:SFM_Spec_Der} is for a finite duration waveform.  For a FM waveform of infinite duration, the $\sinc\left[\pi T \left(f - f_c - f_m n \right) \right]$ function in \eqref{eq:SFM_Spec_Der} is replaced by a delta function expressed as  $\delta\left(f - f_c - f_m n \right)$ resulting in spectral lines of zero width spaced every $f_m$ Hz apart.

\subsection{The GSFM's Spectrum}
\label{subsec:GSFM_SPEC}
For the GSFM waveform, again we use the waveform definition in \eqref{eq:ComplexExpo} over the interval $-T/2 \leq t \leq T/2$ with rectangular window tapering function with height $\frac{1}{\sqrt{T}}$ to ensure unit energy.  The even-symmetric phase GSFM's IF and Phase functions can be represented using a Fourier Series expansion expressed as
\ifthenelse {\boolean{singleColumn}}
{\begin{equation}
f_{GSFM}\left(t\right) = \left(\dfrac{\Delta f}{2}\right) \left[ \frac{a_0}{2} + \sum_{k=1}^{\infty} a_k \cos \left(\dfrac{2\pi k t}{T} \right) \right] + f_c
\label{eq:GSFM_IF_II}
\end{equation}}
{\begin{multline}
f_{GSFM}\left(t\right) = \left(\dfrac{\Delta f}{2}\right) \left[ \frac{a_0}{2} + \sum_{k=1}^{\infty} a_k \cos \left(\dfrac{2\pi k t}{T} \right) \right] + f_c
\label{eq:GSFM_IF_II}
\end{multline}}
where $a_k$ are the Fourier coefficients of $f\left(t\right)$ scaled by the swept bandwidth $\Delta f$ and again the pulse length $T$ defines the period of the Fourier harmonics.  Integrating \eqref{eq:GSFM_IF_II} and multiplying by $2 \pi$ yields the waveform's instantaneous phase
\begin{equation}
\varphi_{GSFM}\left(t\right) = 
\dfrac{\pi \Delta f a_0 t}{2} + \sum_{k=1}^{\infty} \beta_k \sin \left(\frac{2\pi k t}{T}\right) + 2\pi f_c t
\label{eq:GSFM_PHI_II}
\end{equation} 
where $\beta_k = \left(\frac{\Delta f T a_k}{k}\right)$. 
The Fourier transform of the GSFM waveform can now be expressed as 
\ifthenelse {\boolean{singleColumn}}
{\begin{equation}
S_{GSFM}\left(f\right) =\dfrac{1}{\sqrt{T}}\int_{-T/2}^{T/2} \exp\Biggl\{j\sum_{k=1}^{\infty}\beta_k \sin\left(\dfrac{2\pi k t}{T}\right) \Biggr\}  e^{-j2\pi\left(f - f_c - \frac{a_0 \Delta f}{4}\right) t} dt
\label{eq:GSFM_ComplexExpo}
\end{equation}}
{\begin{multline}
S_{GSFM}\left(f\right) =\dfrac{1}{\sqrt{T}} \times \\ \int_{-T/2}^{T/2} \exp\Biggl\{j\sum_{k=1}^{\infty}\beta_k \sin\left(\dfrac{2\pi k t}{T}\right)   \Biggr\}  e^{-j2\pi\left(f - f_c - \frac{a_0 \Delta f}{4}\right) t} dt
\label{eq:GSFM_ComplexExpo}
\end{multline}}
To simplify this expression, we use a Jacobi-Anger-like expression for Generalized Bessel Functions (GBF) \cite{Lorenzutta, Dattoli} expressed as
\ifthenelse {\boolean{singleColumn}}
{\begin{equation}
\exp\Biggl\{j\sum_{k=1}^{\infty}\beta_k \sin\left(\dfrac{2\pi k t}{T}\right)  \Biggr\} = \sum_{n=-\infty}^{\infty}\mathcal{J}_n\{\beta_1,  \beta_2, ..., \beta_{\infty}\}e^{\frac{j2\pi n t}{T}} 
\label{eq:GBF_JAE}
\end{equation}}
{\begin{equation}
\exp\Biggl\{j\sum_{k=1}^{\infty}\beta_k \sin\left(\dfrac{2\pi k t}{T}\right)  \Biggr\} = \sum_{n=-\infty}^{\infty}\mathcal{J}_n^{1:\infty}\{\beta_k\}e^{\frac{j2\pi n t}{T}} 
\label{eq:GBF_JAE}
\end{equation}}
where $\mathcal{J}_n^{1:\infty}\{\beta_k\} = \mathcal{J}_n\{\beta_1,  \beta_2, ..., \beta_{\infty}\}$ is the infinite-dimensional GBF of the first kind.  The GBF is generalized in that it accepts an infinite-dimensional argument.  Using \eqref{eq:GBF_JAE}, the integral in \eqref{eq:GSFM_ComplexExpo} simplifies to
\ifthenelse {\boolean{singleColumn}}
{\begin{equation}
S_{GSFM}\left(f\right) = \sum_{n=-\infty}^{\infty}\mathcal{J}_n^{1:\infty}\{\beta_k\}  \int_{-T/2}^{T/2}e^{-j2\pi\left(f - f_c - \frac{a_0 \Delta f }{4} - \frac{n}{T} \right) t}dt
\label{eq:GSFM_Spec_Integral}
\end{equation}}
{\begin{multline}
S_{GSFM}\left(f\right) = \sum_{n=-\infty}^{\infty}\mathcal{J}_n^{1:\infty}\{\beta_k\} \times \\ \int_{-T/2}^{T/2}e^{-j2\pi\left(f - f_c - \frac{a_0 \Delta f }{4} - \frac{n}{T} \right) t}dt
\label{eq:GSFM_Spec_Integral}
\end{multline}}
Finally, evaluating the integral in \eqref{eq:GSFM_Spec_Integral} yields the result
\ifthenelse {\boolean{singleColumn}}
{\begin{equation}
S_{GSFM}\left(f\right) = \sqrt{T}\sum_{n=-\infty}^{\infty}\mathcal{J}_n^{1:\infty}\{\beta_k\}  \sinc\left[\pi T \left(f - f_c - \frac{a_0 \Delta f}{4} - \frac{n}{T} \right)\right]
\label{eq:GSFM_Spec_Der_II}
\end{equation}}
{\begin{multline}
S_{GSFM}\left(f\right) = \sqrt{T}\sum_{n=-\infty}^{\infty}\mathcal{J}_n^{1:\infty}\{\beta_k\} \times \\  \sinc\left[\pi T \left(f - f_c - \frac{a_0 \Delta f}{4} - \frac{n}{T} \right)\right]
\label{eq:GSFM_Spec_Der_II}
\end{multline}}
Reassuringly, for the case where $\rho = 1.0$ (i.e., an SFM waveform), the expression \eqref{eq:GSFM_Spec_Der_II} collapses back to the spectrum of the SFM waveform as shown in \eqref{eq:SFM_Spec_Der}.  For the case of a SFM waveform the Fourier Series for the SFM's IF function contains only the single harmonic $f_m$.  The single Fourier coefficient is 
\begin{equation}  \beta_k = \left\{
\begin{array}{ll}
      \frac{\Delta f}{2 f_m}, & k = 1 \\
      
      0, & otherwise \\
\end{array} 
\right.
\label{eq:SFM_Fourier_Series} 
\end{equation}
By utilizing the reduced dimension identity of the GBF's \cite{Lorenzutta}
\begin{equation}
\mathcal{J}_n\{x, 0, ..., 0\} = J_n\{x\}
\label{eq:GBF_Identity}
\end{equation}
the GBF's become the one-dimensional cylindrical Bessel function of the first kind.  Utilizing the identity in \eqref{eq:GBF_Identity} and setting Fourier harmonic period to $1/f_m$, the expression in \eqref{eq:GSFM_Spec_Der_II} collapses back into the expression for the SFM's spectrum \eqref{eq:SFM_Spec_Der}.

\subsection{Derivation of Carson's Bandwidth Rule for the GSFM Waveform}
\label{subsec:GSFM_Carson}
In deriving Carson's bandwidth rule for the GSFM, we exploit the GSFM phase and IF functions defined in \eqref{eq:GSFM_PHI} and \eqref{eq:GSFM_IF} respectively.  Carson's bandwidth rule \cite{Couch} states that $98\%$ of a FM waveform's energy resides in a bandwidth $B$ expressed as 
\begin{equation}
B = 2\left(\beta + 1\right)B_m = \Delta f + 2 B_m
\label{eq:Carson}
\end{equation}
where $\Delta f$ is the FM waveform's swept bandwidth, $B_m$ is the highest frequency component of the waveform's IF function, and $\beta = \Delta f / 2B_m$ is the Frequency Deviation Ratio (FDR) \cite{Couch}.  To find $B_m$ for the GSFM waveform, we need to find the highest frequency component present in \eqref{eq:GSFM_IF} which for the non-symmetric IF GSFM occurs at time $t = T$, where $T$ is the duration of the waveform.  Finding $B_m$ requires deriving the IF function of the GSFM's IF function \eqref{eq:GSFM_IF}, denoted as $f_{IF}\left(t\right)$, which is expressed as  
\begin{equation}
f_{IF}\left(t\right) = \alpha \rho t^{\left(\rho-1\right)}
\label{eq:IFIF}
\end{equation} 
Evaluating \eqref{eq:IFIF} at $t = T$ yields the result
\begin{equation}
B_m = \alpha \rho T^{\left(\rho-1\right)}
\label{eq:B_m1}
\end{equation}    
Applying the result in \eqref{eq:B_m1} to \eqref{eq:Carson} results in a $98\%$ bandwidth of 
\ifthenelse {\boolean{singleColumn}}
{\begin{equation}
B_{GSFM} = 2\left(\dfrac{\Delta f}{2\alpha \rho T^{\left(\rho - 1\right)}} + 1\right)\alpha \rho T^{\left(\rho - 1\right)} = \Delta f + 2\alpha \rho T^{\left(\rho - 1\right)}.
\label{eq:GSFM_Carson_I}
\end{equation}}
{\begin{equation}
B_{GSFM} = \Delta f + 2\alpha \rho T^{\left(\rho - 1\right)}.
\label{eq:GSFM_Carson_I}
\end{equation}}
Note that for the case when $\rho = 1$ (i.e. an SFM), $\alpha$ becomes the SFM's modulation frequency $f_m$ and \eqref{eq:GSFM_Carson_I} becomes 
$2\left(\beta + 1\right)f_m$, Carson's bandwidth rule for the SFM waveform. 

\subsection{The GSFM's PAPR}
\label{subsec:GSFM_PAPR}
Using \eqref{eq:ComplexExpo} with unit amplitude and utilizing the Jacobi-Anger expansion for GBFs, the complex analytic waveform is 
\begin{equation}
s\left(t\right) = \sum_{m=-\infty}^{\infty} \mathcal{J}_m^{1:\infty}\{\beta_k\} e^{j2\pi\left[\left(\frac{m}{T}\right) + \left(f_c+ \frac{\Delta f a_0}{4}\right)\right]t}.
\end{equation}
The real component $x\left(t\right)$ is
\ifthenelse {\boolean{singleColumn}}
{\begin{equation}
x\left(t\right) = \sum_{m=-\infty}^{\infty} \mathcal{J}_m^{1:\infty}\{\beta_k\} \cos\left[2\pi\left(\left(\frac{m}{T}\right) + \left(f_c+ \frac{\Delta f a_0}{4}\right)\right)t\right].
\label{eq:realJAE}
\end{equation}}
{\begin{multline}
x\left(t\right) = \sum_{m=-\infty}^{\infty} \mathcal{J}_m^{1:\infty}\{\beta_k\} \times \\ \cos\left[2\pi\left(\left(\frac{m}{T}\right) + \left(f_c+ \frac{\Delta f a_0}{4}\right)\right)t\right].
\label{eq:realJAE}
\end{multline}}
We assume that the carrier terms result in a sinusoid with an integer number of cycles $L = \left(f_c+ \frac{\Delta f a_0}{4}\right)T$ in the waveform's duration.  Additionally, we make the approximation that the vast majority of the waveform's energy is contained in the central $2M + 1$ terms of the summation in \eqref{eq:realJAE} with $M < L$ resulting in the expression
\begin{equation}
x\left(t\right) \cong \sum_{m=-M}^{M} \mathcal{J}_m^{1:\infty}\{\beta_k\} \cos\left[2\pi\left(\frac{m + L}{T}\right)t\right].
\label{eq:truncate}
\end{equation}
These assumptions ensure that the spectral extent of the waveform $\frac{2M+1}{L}$ is not more than twice the center frequency $\left(f_c+ \frac{\Delta f a_0}{4}\right)$ preventing any non-zero spectral content at DC thus maintaining the mathematical requirements for the complex analytic signal model.  The truncated summation \eqref{eq:truncate} implies that the GBF $\mathcal{J}_m\{\beta_k\}$ is negligibly small for order $m > M$, which is a general property of GBFs \cite{Dattoli}.

The magnitude-squared of the real-valued waveform times series
\begin{multline}
|x\left(t\right)|^2 \cong \sum_{m,n=-M}^{M} \mathcal{J}_m^{1:\infty}\{\beta_k\}\mathcal{J}_n^{1:\infty}\{\beta_k\} \times \\ \cos\left[2\pi\left(\frac{m + L}{T}\right)t\right]\cos\left[2\pi\left(\frac{n + L}{T}\right)t\right].
\label{eq:maxSquare}
\end{multline}
Using the definition of the PAPR in \eqref{PAPR}, $\max_t\{|x\left(t\right)|^2\} = 1$.  The average power, the denominator of \eqref{PAPR}, is
\ifthenelse {\boolean{singleColumn}}
{\begin{multline}
\frac{1}{T}\int_{-T/2}^{T/2}|x\left(t\right)|^2dt \cong \sum_{m=-M}^{M}\sum_{n=-M}^{M} \mathcal{J}_m^{1:\infty}\{\beta_k\}\mathcal{J}_n^{1:\infty}\{\beta_k\} \times \\ \frac{1}{2T}\int_{-T/2}^{T/2} \cos\left[2\pi\left(\frac{m + L}{T}\right)t\right]\cos\left[2\pi\left(\frac{n + L}{T}\right)t\right]dt.
\label{eq:rms1}
\end{multline}}
{\begin{multline}
\frac{1}{T}\int_{-T/2}^{T/2}|x\left(t\right)|^2dt \cong \sum_{m,n=-M}^{M} \mathcal{J}_m^{1:\infty}\{\beta_k\}\mathcal{J}_n^{1:\infty}\{\beta_k\} \times \\ \int_{-T/2}^{T/2} \cos\left[2\pi\left(\frac{m + L}{T}\right)t\right]\cos\left[2\pi\left(\frac{n + L}{T}\right)t\right]dt.
\label{eq:rms1}
\end{multline}}
Since the cosine components are orthogonal to each other, the expression in \eqref{eq:rms1} can be simplified to 
\begin{multline}
\frac{1}{T}\int_{-T/2}^{T/2}|x\left(t\right)|^2dt \cong \sum_{m=-M}^{M}\frac{1}{T} \left(\mathcal{J}_m^{1:\infty}\{\beta_k\}\right)^2 \times \\ \int_{-T/2}^{T/2} \cos^2\left[2\pi\left(\frac{m + L}{T}\right)t\right]dt \\ = \frac{1}{2T}\sum_{m=-M}^{M} \left(\mathcal{J}_m^{1:\infty}\{\beta_k\}\right)^2 \times \\ \int_{-T/2}^{T/2}\left(1+\cos\left[4\pi\left(\frac{m + L}{T}\right)t\right]\right)dt .
\label{eq:maxSquareSimp}
\end{multline}
Evaluating the integral yields the result
\begin{multline}
\frac{1}{T}\int_{-T/2}^{T/2}|x\left(t\right)|^2dt \cong \frac{1}{2T}\sum_{m=-M}^{M} \left(\mathcal{J}_m^{1:\infty}\{\beta_k\}\right)^2 \times \\ \int_{-T/2}^{T/2}\left(1+\cos\left[4\pi\left(\frac{m + L}{T}\right)t\right]\right)dt  \\ = \sum_{m=-M}^{M} \left(\mathcal{J}_m^{1:\infty}\{\beta_k\}\right)^2 \left[\frac{1 + \sinc\left(2\pi\left(m+L\right)\right)}{2}\right].
\label{eq:rms2}
\end{multline}
Since $m$ and $L$ are integers and the minimum value of $m$ is $-M$ where $M < L$, the sinc function will always evaluate to zero.  This simplifies \eqref{eq:rms2} to 
\begin{equation}
\frac{1}{T}\int_{-T/2}^{T/2}|x\left(t\right)|^2dt \cong \\ \left(\frac{1}{2}\right) \sum_{m=-M}^{M} \left(\mathcal{J}_m^{1:\infty}\{\beta_k\}\right)^2.
\label{eq:rms3}
\end{equation}
As noted in \cite{Dattoli}, $\sum_{m=-\infty}^{\infty} \left(\mathcal{J}_m^{1:\infty}\{\beta_k\}\right)^2 = 1$.  Recalling that the approximation for $x\left(t\right)$ \eqref{eq:truncate} implied that $\mathcal{J}_m^{1:\infty}\{\beta\}$ is negligibly small for $m > M$, this identity holds approximately and thus simplifying \eqref{eq:rms3} to $1/2$.  The PAPR in linear scale can now be stated as
\begin{equation}
\PAPR = \left(\dfrac{\max_t\{|x\left(t\right)|^2\}}{\frac{1}{T}\int_0^T|x\left(t\right)|^2dt}\right) \cong \frac{1}{\left(1/2\right)} = 2
\end{equation}
or $10 \log_{10}\left(2\right) = 3.02$ dB, what one would expect from a constant amplitude CW or LFM waveform.
\begin{acknowledgments}
This research was supported in part by ONR Grants N00014-12-1-0047 and N00014-15-1-2238.  David A. Hague was funded by the SMART Scholarship program.  
\end{acknowledgments}

\end{document}